%% file: TPhi.tex
\documentclass[prd,twocolumn,amsmath,amssymb,floatfix,nofootinbib,superscriptaddress]{revtex4}

\usepackage{graphicx}
\usepackage{bm}
\usepackage{amsmath,amssymb}
\usepackage{epsfig}
\usepackage{color}
\usepackage{natbib}
\usepackage{enumerate}
\usepackage[hypertex]{hyperref}
\usepackage{ifthen}
\usepackage{subfigure}

\input{macros.tex}

\newcommand{\tC}{\tilde{C}}

\newcommand{\tT}{\tilde{T}}

\def\eprinttmp@#1arXiv:#2 [#3]#4@{
\ifthenelse{\equal{#3}{x}}{\href{http://arxiv.org/abs/#1}{#1}}{\href{http://arxiv.org/abs/#2}{arXiv:#2} [#3]}}

\providecommand{\eprint}[1]{\eprinttmp@#1arXiv: [x]@}
\newcommand{\adsurl}[1]{\href{#1}{ADS}}
\providecommand{\bibinfo}[2]{\ifthenelse{\equal{#1}{isbn}}{
\href{http://cosmologist.info/ISBN/#2}{#2}}{#2}}

\newcommand{\temp}{T}

\newcommand{\eff}{{\rm{eff}}}

\begin{document}


\title{The full squeezed CMB bispectrum from inflation}

\author{Antony Lewis}
\homepage{http://cosmologist.info}
\address{Department of Physics \& Astronomy, University of Sussex, Brighton BN1 9QH, UK}

\begin{abstract}
The small-scale CMB temperature we observe on the sky is modulated by perturbations that were super-horizon at recombination, giving differential focussing and lensing that generate a non-zero bispectrum even for single-field inflation where local physics is identical. Understanding this signal is important for primordial non-Gaussianity studies and also parameter constraints from the CMB lensing bispectrum signal. Because of cancellations individual effects can appear larger or smaller than they are in total, so a full analysis may be required to avoid biases. I relate angular scales on the sky to physical scales at recombination using the optical equations, and give full-sky results for the large-scale adiabatic temperature bispectrum from Ricci focussing (expansion of the ray bundle), Weyl lensing (convergence and shear), and temperature redshift modulations of small-scale power. The $\delta N$ expansion of the beam is described by the constant temperature 3-curvature,
and gives a nearly-observable version of the consistency relation prediction from single-field inflation. I give approximate arguments to quantify the likely importance of dynamical effects, and argue that they can be neglected for modulation scales $l\alt 100$, which is sufficient for lensing studies and also allows robust tests of local primordial non-Gaussianity using only the large-scale modulation modes. For accurate numerical results early and late-time ISW effects must be accounted for, though I confirm that the late-time non-linear Rees-Sciama contribution is negligible compared to other more important complications. The total corresponds to $\fnl \sim 7$ for Planck-like temperature constraints and $\fnl \sim 11$ for cosmic-variance limited data to $\lmax=2000$. Temperature lensing bispectrum estimates are affected at the $0.2\sigma$ level by Ricci focussing, and up to $0.5\sigma$ with polarization.
\end{abstract}

\date{\today}

\maketitle

\pagenumbering{arabic}

\section{Introduction}

\begin{figure}
\includegraphics[width=8cm]{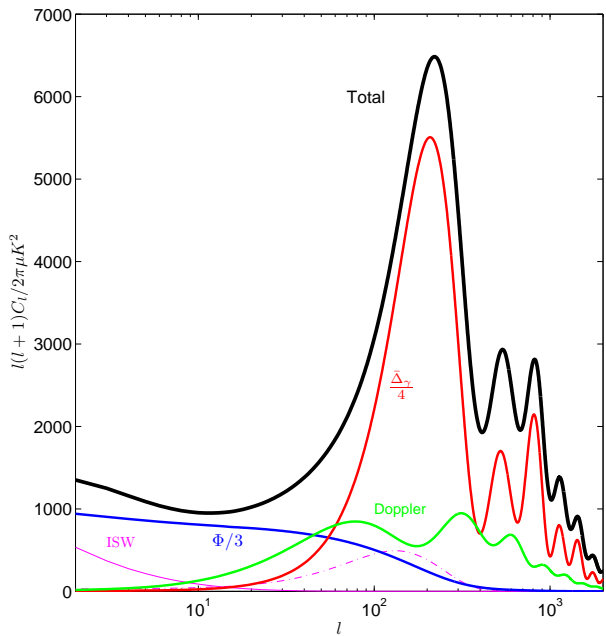}
\caption{
Power spectrum of the contributions to the total CMB temperature anisotropy $C_l$. The red $\bar{\Delta}_\gamma$ contribution is the main small-scale contribution from the comoving temperature perturbations at last-scattering, with $\Phi/3$ being the net large-scale contribution from photons climbing out of potential wells (assuming matter domination).   Note that the $\Phi/3$ and $\bar{\Delta}_\gamma$ source terms have opposite sign, so their total contribution to the power spectrum is nearly zero at $l\sim 60$, where the total is then dominated by the Doppler term (green). The magenta ISW contributions come from the late-time change in the potentials when dark energy becomes important at low redshift (solid), and the early contribution (dash-dotted) from time-varying potentials soon after recombination as the universe became fully matter rather than radiation dominated.
\label{TTcontribs}
}
\end{figure}

Primordial local non-Gaussianity can provide strong constraints on early-universe physics, and the lensing bispectrum can be used to constrain dark energy models.
A non-zero CMB bispectrum should be detected by the Planck satellite, and it is important to understand the physical origins of this signal.

In principle a non-linear perturbation-theory numerical calculation should be able to calculate the expected signal (e.g. as first attempted by ~\cite{Pitrou:2010sn}), but such a
calculation is both conceptually and numerically complex, and as yet has not been done convincingly. Fortunately however, both lensing and local non-Gaussianity signals
are dominated by squeezed bispectrum shapes. These correspond to large-scale modulations of the small-scale power, where the modulating field is correlated to the large-scale CMB temperature anisotropies. In the limit in which modulating adiabatic curvature perturbations are super-horizon at recombination, the local physics in each Hubble patch should be identical, and in this case analytic arguments can be used to estimate the bispectrum~\cite{Creminelli:2004pv,Boubekeur:2009uk,Bartolo:2011wb,Creminelli:2011sq}. In this paper I re-derive previous results by directly relating observed angles on the sky to physical scales at recombination. This gives results that are valid in a more weakly squeezed approximation, include linear transfer and sky curvature effects, and hence are valid over a wider range of angular scales so that they can be directly applied to data. Including all the relevant effects may be important because of partial cancellations, especially for understanding the lensing signal. I also discuss a detailed numerical calculation of the dominating lensing part of the signal, and estimate the scale at which complications due to sub-horizon dynamics are likely to become important.

Before getting into a discussion of the bispectrum, it is worth briefly remembering the various effects that give rise to the observed linear CMB temperature anisotropies. Recombination is an approximately equal-temperature surface, so observers close to the surface would see the same thing everywhere if there were no sub-horizon perturbations. However, we see temperature anisotropies on all scales because photons from different parts of the surface get redshifted by differing amounts. On scales that are super-horizon at recombination the photon redshifting effects are dominated by the potential-well Sachs-Wolfe effect, and the late-time Integrated Sachs-Wolfe Effect (ISW) from passing through evolving potential wells along the line of sight. On sub-horizon scales overdensities recombine later and are redshifted less and hence appear hot, and at around $l\sim 60$ this contribution cancels with the Sachs-Wolfe, so that the anisotropies are actually dominated by Doppler signals from velocities at last-scattering. As shown in Fig.~\ref{TTcontribs} there is no scale on which the Sachs-Wolfe limit is accurate, and only at $l\ll 60$ are Doppler effects negligible. In the region $10\alt l \alt 100$ the signal has contributions from Doppler, Sachs Wolfe and density perturbations of comparable magnitudes, as well as a significant early Sachs-Wolfe contribution. For a squeezed-limit result to be trustworthy over a useful range of scales it should account for these various effects and be not be too sensitive to the presence of the first corrections from sub-horizon dynamical evolution.

The large-scale temperature anisotropies are generated by large-scale perturbations modulating the amount of redshifting of a uniform-temperature last-scattering surface. The bispectrum arises from the small-scale \emph{anisotropies} on this surface being modulated by the large-scale perturbations. If there are only super-horizon modes from single-field inflation present, the physics at recombination should be identical in different Hubble patches, and the small-scale anisotropies would locally look statistically equivalent. However when we observe the surface these small-scale anisotropies can be modulated by large-scale perturbations that change the amount of Ricci and Weyl focussing, Weyl shearing, and redshifting that we see in different parts of the sky. Since these modulating perturbations also generate the large-scale CMB temperature anisotropies, the modulation is correlated to the temperature and there is a bispectrum.

I calculate this signal by integrating along a ray bundle to relate observed angles to physical scales on the last-scattering surface; a concrete calculation of the `projection effects' discussed by Ref.~\cite{Senatore:2012nq}. This gives a simple derivation of results that are valid on the full sky (no flat-sky approximation, which is important for large-scale modulations), and results can easily incorporate the linear-theory transfer functions to include all the relevant terms in the modulation correlation function. Previous work has either focussed on the CMB lensing bispectrum alone, or has done a full approximate flat-sky calculation neglecting the part of the lensing bispectrum that is not due to super-horizon modes. However this separation is not quantitatively justified, since the lensing bispectrum has significant contributions from early and late times (from early and late ISW effects), as well as large scale modes and other sources. The sources do not separate cleanly in either redshift or wavenumber, so it is best to do a full self-consistent analysis.

Previous work has emphasised that there are partial cancellations between Ricci focussing and the lensing (non-ISW) Weyl focussing terms~\cite{Creminelli:2004pv,Boubekeur:2009uk,Creminelli:2011sq}, and a full analysis is therefore potentially important for lensing bispectrum studies (as well for a consistent bispectrum calculation for subtraction when looking for primordial non-Gaussianity). This cancellation is very similar to that in the linear CMB anisotropies, where the total large-scale anisotropy is smaller than individual terms (see Appendix~\ref{linearCMB}).

The Ricci focussing modulation accounts for the well-known `consistency relation' result for the bispectrum in single-field inflation~\cite{Maldacena:2002vr,Creminelli:2011sq}: calculated here in the context of the directly-observable CMB, which has acoustic oscillations, the signal is rather larger than it appears in the primordial curvature perturbations (which have a very smooth spectrum so the signal is completely negligible).

In practice the (Weyl) lensing signal is often most easily calculated by reconstructing the lensing potential using quadratic maximum likelihood estimators~\cite{Okamoto03} (see Ref.~\cite{Hanson:2009kr} for a review), and then correlating it with the temperature to get the bispectrum: in the squeezed limit the temperature-lensing reconstruction power spectrum contains the same information as the bispectrum. By using a lensing potential, shear and convergence effects are treated self-consistently in one go since they are related by gradients at linear order for scalar perturbations. For this reason I will group the lensing effects separately from the others, and quantify the importance of the partial cancellations for the full result. This allows biases in a pure-lensing analysis to be assessed and corrected, and to quantify whether the other terms are significant. My classification is therefore closer to Ref.~\cite{Bartolo:2011wb} --- who exclude Weyl lensing altogether --- than Ref.~\cite{Creminelli:2011sq}, who include early Weyl lensing in their combined result but don't include the larger late-time signal. The full result can easily be obtained by adding the terms together. I consider the full result including polarization in the final section.

The approach followed here is similar to the analysis of redshift-space densities in Refs.~\cite{Challinor:2011bk}. Ref.~\cite{Schmidt:2012ne} also recently developed general results for distortions to standard rulers in redshift space. The main difference for the CMB is that we are now concerned with a single constant temperature recombination surface (rather than constant redshift surfaces), and that the CMB is at much higher redshift so the line-of-sight impact of super-horizon perturbations is relatively more important.

\subsection{Jacobi map between physical scales and observed angles}


\begin{figure*}
\includegraphics[width=8.5cm]{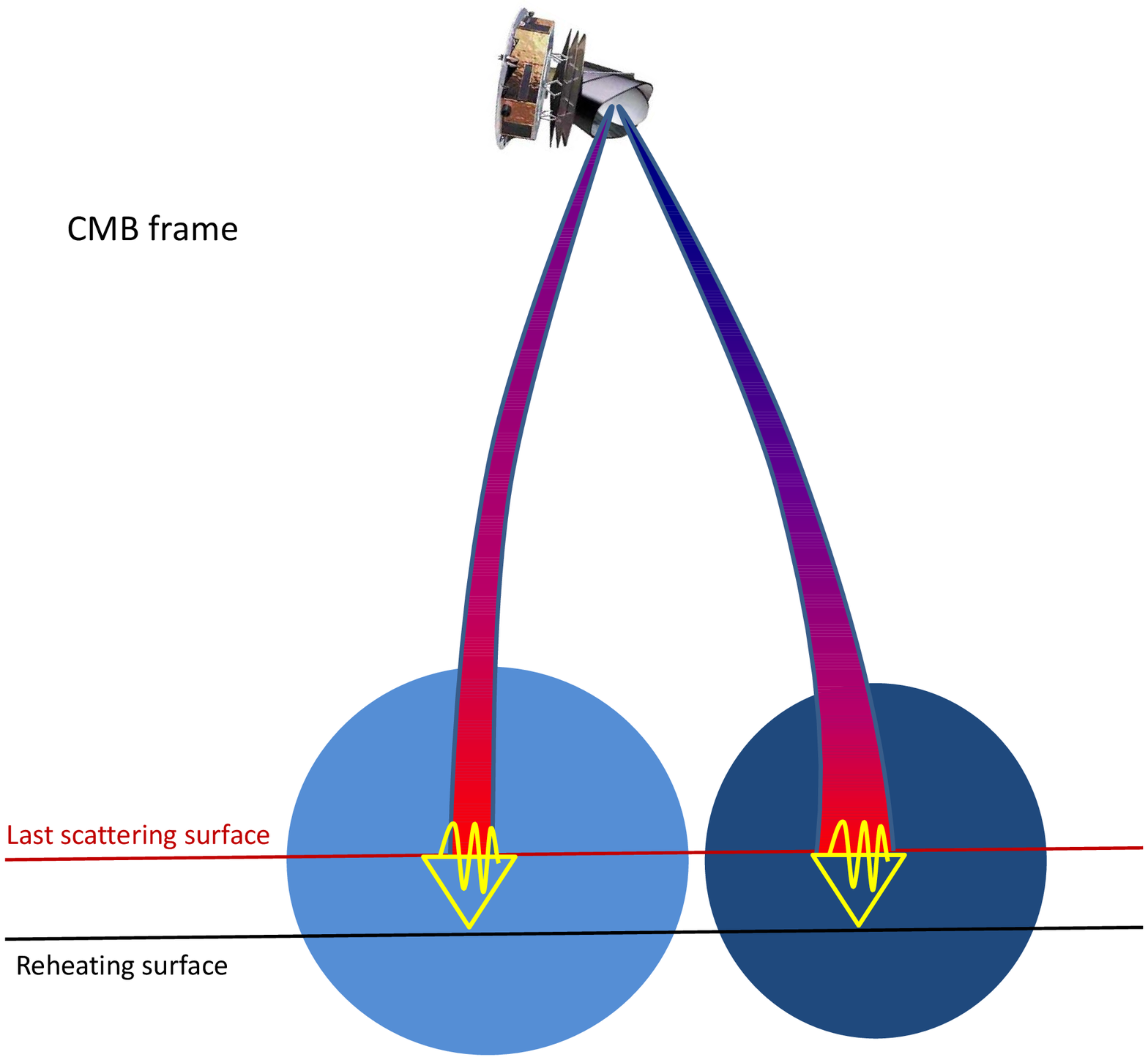}
\includegraphics[width=8.5cm]{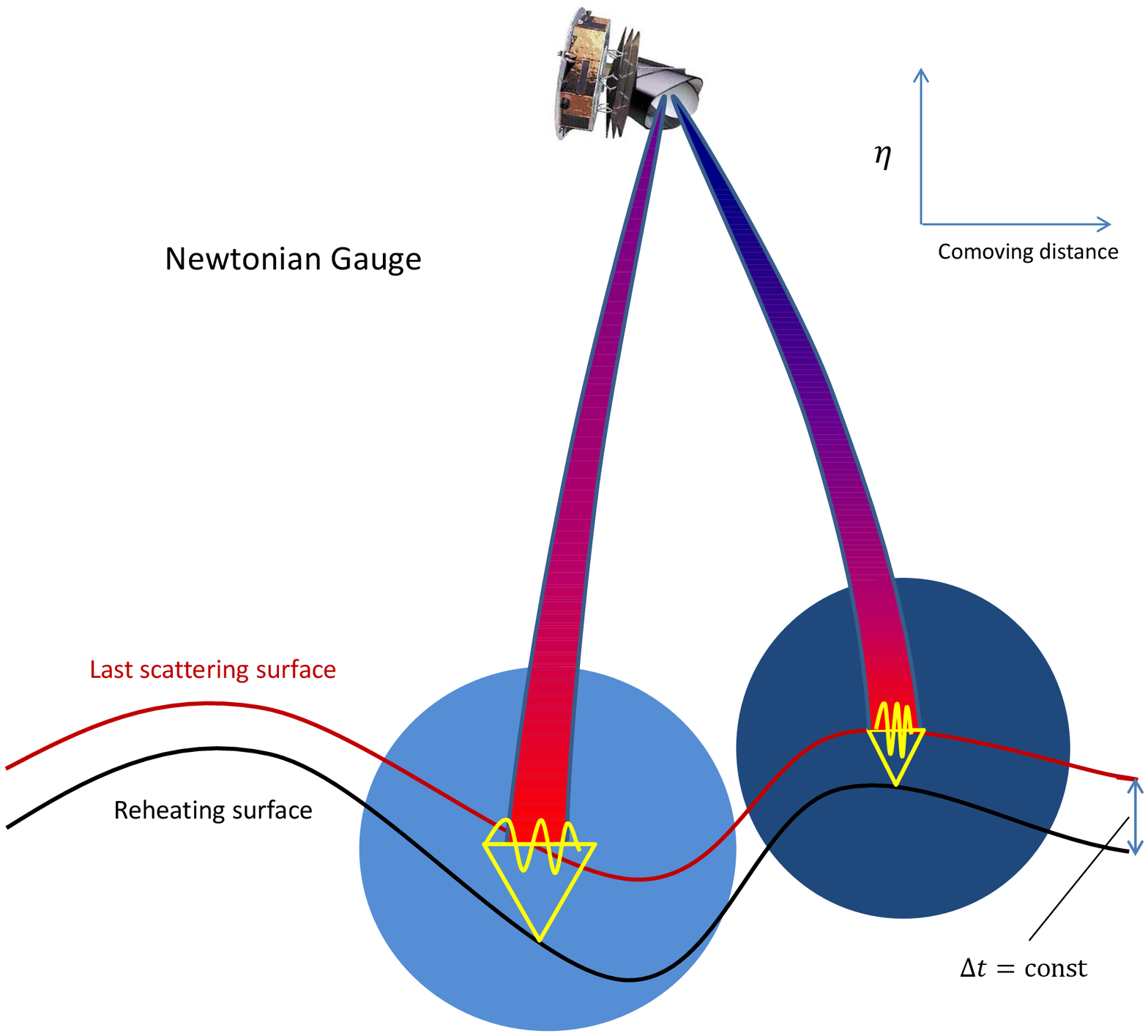}
\caption{
Schematic equivalent descriptions of an observation of fixed physical-size small-scale perturbations on the equal-temperature recombination surface, in the presence of large perturbations that were super-horizon at recombination.  The perturbation on the right side of each figure appears to have slightly smaller size on the sky. Ricci (de)focussing directly changes the cross-section of the ray bundle due to the expansion of spacetime along the beam. Weyl lensing bends the path of the beam, causing Weyl focussing (and shear) by differential deflection. Angular anisotropy in the Ricci focussing is described gauge-invariantly by the constant temperature 3-curvature. In the left figure, a non-comoving gauge with constant temperature slices, Ricci (de)focussing is what changes the observed angular size of the perturbations. In the comoving Newtonian gauge on the right figure, the effect can be described as the overdensity having recombination happening later, so fixed physical scales have smaller comoving size and appear smaller in the background, combined with a perturbation to the scale factor at the unperturbed background comoving distance described by the potential.
The Ricci focussing effect that makes the patch look smaller is then partly compensated by Weyl focussing due to differential deflection of the beams in the potential gradients along the line of sight; this cancellation is not shown to more clearly illustrate the Ricci effect. The right patch also looks colder because there is more redshifting from the photons climbing out of the deeper potential well than the partly-compensating blue-shifting from the beam contraction (see Appendix~\ref{linearCMB}; note sub-horizon overdensities would look hotter).
\label{diagram}
}
\end{figure*}

\begin{figure}
\includegraphics[width=8.5cm]{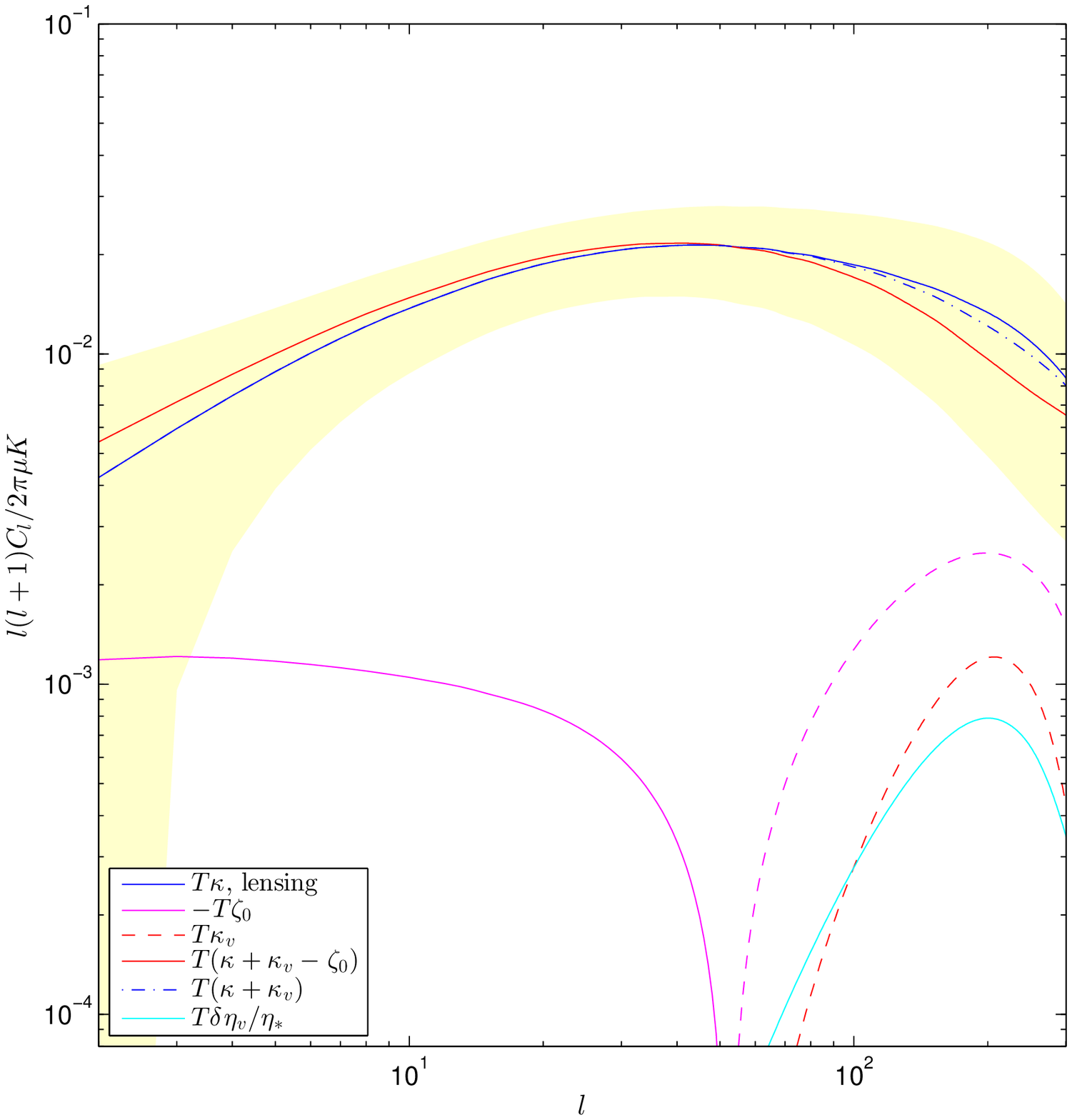}
\caption{The CMB temperature cross-correlation power spectra relevant for contributions to the bispectrum (dashed lines are negative).
The blue curve is the `standard' result for lensing convergence and shear, magenta shows the contribution from
Ricci focussing determined by the super-horizon curvature perturbation $\zeta_0^*$ without evolution (the correlation changes sign at $l\sim 60$ where Sachs-Wolfe and comoving overdensity terms cancel in the temperature). At $l\agt 100$ sub-horizon effects may become comparable, with terms estimated for  large-scale flows (red dashed) and perturbations in sound-horizon (cyan).
The solid red curve shows an estimate of the total effective magnification correlation including flows and Ricci focussing.
The total effective shear is only modified by sub-horizon dynamics, and is estimated by the dash-dot blue line. The sub-horizon effects are unlikely to be accurate, but give an indication of the scale on which they might become relevant.
The shaded band shows an indicator of $\pm1\sigma$ sensitivity to systematic errors relative to the cosmic variance limit from lensing, given by $\pm \sqrt{ C_l^{TT} C^{\kappa\kappa}_l + (C_l^{T\kappa})^2}/l$.
\label{correlations}
}
\end{figure}

For single-field inflationary fluctuations, there is only one clock (e.g. set by the temperature), so locally physics is the same at recombination in all patches that differ only by super-horizon modes (up to gradients that we consider separately later)~\cite{Maldacena:2002vr,Creminelli:2004yq,Senatore:2012nq}. However, the small-scale physical scales at recombination are observed at different angular sizes and temperatures because of the large-scale perturbations. Differences in redshifting from the uniform temperature surface in different directions gives rise to the linear CMB anisotropies (see Appendix~\ref{linearCMB} for a brief review). Similarly the same physical sizes appear at different angular sizes in different directions, due to bending of the photon path and differences in expansion along the line of sight. The path deflections also give rise to shear distortions.

I calculate the anisotropy in observed angular sizes and shear by using the optical equations for the propagation of a ray bundle through the perturbed universe. For a pair of rays separated by an infinitesimal angular coordinate $\delta\theta_J$ about an observation direction $\vnhat$ at $A$, the physical infinitesimal separation vector orthogonal to the direction of propagation $\xi_I$ at affine parameter $\lambda$ along the ray is given by
 \be
 \xi_I(\lambda) = \cld_{IJ}(\lambda)\delta\theta_J,
 \ee
 where $\cld_{IJ}$ is called the Jacobi map (see Ref.~\cite{Lewis:2006fu} for detailed review in similar notation). In general $\cld_{IJ}$ can be found by integrating the optical equations along the line of sight~\cite{Sachs61}. The trace of $\cld_{IJ}$ determines the (linearized) magnification and the symmetric trace-free part determines the shear. Scalar perturbations generate no rotation at first order so the matrix is symmetric.

For simplicity I restrict to a flat, almost-Friedmann-Robertson-Walker (FRW) model with scalar
perturbations,  working to first order in the conformal-Newtonian gauge (CNG) with metric
\begin{equation}
d s^2 = a^2(\eta) [(1+2\Psi)d\eta^2 - (1-2\Phi) \delta_{ij} d x^i d x^j].
\label{eq:3}
\end{equation}
 The trace of the Jacobi map $\cld$ depends on the Ricci tensor and determines the observed angular size; it gives a generalization of the angular diameter distance to a perturbed universe\footnote{This assumes the source plane is ray-orthogonal, however the correction is next order so we can neglect it.} (see e.g. Refs.~\cite{Lewis:2006fu,Bernardeau:2009bm,Challinor:2011bk})
\be
\cld(\hat{\vn},\eta)/2 = \chi(\hat{\vn},\eta) a(\eta)[1+\Phi_A-\Phi -\kappa + \vnhat\cdot \vv_A].
\label{cld}
\ee
Here the lensing convergence $\kappa$ is determined solely by the Weyl focussing along the line of sight, $\vv_A$ is the Newtonian-gauge (peculiar) velocity of the observer at $A$, and the local potential $\Phi_A$ is included so that the observer's scale factor is defined to be unity, $a(0)(1-\Phi_A)=1$. The potential at the source $\Phi$ looks like a local term, but is just the result of integrating a total derivative part of the Ricci tensor along the ray: it accounts for the line-of-sight Ricci focussing that is not accounted for in the convergence term $\kappa$ generated by Weyl focussing.

The Riemann tensor only depends on second derivatives of $\Phi$ and $\Psi$, so only these gradients are observable: a constant $\Phi$ cancels in Eq.~\eqref{cld} so only differences in potentials are observable. A constant $\vgrad^2\Phi$ gives a physical monopole contribution to $\cld$ but is equivalent to the perturbative effect of a non-flat FRW geometry~\cite{Bernardeau:2009bm}, so I shall not consider the monopole further and drop $\Phi_A$. As is usual in CMB lensing studies we can then define a lensing potential for a source at radial distance $\chi_*$
\be
\psi \equiv -2\int_0^{\chi_*} \d\chi \frac{\chi_*-\chi}{\chi\chi_*} \Psi_W(\chi\vnhat,\eta_A-\chi)
\ee
where the Weyl potential $\Psi_W\equiv (\Phi+\Psi)/2$ determines the scalar Weyl (conformally invariant) part of the Riemann tensor, so the deflection angle is $\grad_I\psi$ and the lensing convergence and shear are given by $\kappa=-\grad_I\grad^I\psi/2$, $\gamma_{IJ} = \grad_{\la I}\grad_{J\ra}\psi$. I choose to define `lensing' to be the Weyl effects, not including the Ricci focussing term that only depends on differences in potentials.


 For modulation scales that are large compared to the thickness of the last-scattering surface it is a good approximation to treat the recombination visibility as a delta function at time $\eta_*$, a distance $\chi_*$ away in an FRW universe. Recombination happens when the universe cools enough so that hydrogen can recombine, so in this approximation last-scattering is an equal-temperature surface. We can then consider how small-scale sub-horizon perturbations on this approximately equal temperature surface are observed, allowing for the fact that there are large-scale perturbations that do not change the local physics at the source --- see Fig.~\ref{diagram}.
 If $\eta_*$ is the background (conformal) time of recombination, in the perturbed universe recombination will happen at a perturbed conformal time
  $\eta=\eta_*+\delta\eta$ (on sub-horizon scales this is the physical effect that overdensities need longer to expand to the recombination temperature): for an equal temperature surface with fixed photon energy density we need $\rho_\gamma'\delta\eta = -\delta\rho_\gamma$ and hence
  $\delta a/a=\clh\delta\eta = \frac{\Delta_\gamma}{4}$, where $\Delta_\gamma$ is the Newtonian-gauge fractional photon density perturbation. In terms of the background scale factor $a_*$ and radial distance $\chi_*$, the physical scales on an equal temperature surface are then related to angles to the sky by
\be
\cld/2 = \chi_*a_*\left[1 + \frac{\delta\chi}{\chi_*} + \frac{\Delta_\gamma}{4} -\Phi -\kappa + \vnhat\cdot\vv_A\right].
\label{fulljabobi}
\ee
Recall that recombination happens at a time $\eta_*\sim 270\rm{Mpc}$ and the distance to last scattering in a standard \LCDM\ model is $\chi_*\sim 14\rm{Gpc}$, so $\eta_*/\chi_*\sim \clo(0.01)$. The $\clo(\Delta_\gamma\eta_*/\chi_*)$ term in the radial displacement $\delta\chi/\chi_*$ is therefore much smaller than the others and can be neglected, and the time delay contribution is also very small due to cancellations~\cite{Hu:2001yq}.
The effect of lensing convergence dominates because although the potentials are roughly constant after recombination the signal is enhanced (even incoherently) by the large number $\clo(\chi_*/\eta_*)$ of perturbations along the line of sight; the convergence signal is also blue because of the two angular gradients. However, there is only a \emph{correlation} of the convergence with the CMB temperature (and hence a bispectrum) from very-large scale lenses, lenses close to the last scattering surface, and the integrated Sachs-Wolfe effect (see later section for detailed discussion).  The term $\frac{\Delta_\gamma}{4} -\Phi$ is of order $\clh\delta\eta$ and significantly smaller, but highly correlated to the temperature and potentially marginally detectable in the bispectrum (and hence also potentially leading to biases if neglected): one of the aims of this paper is to quantify more carefully the importance of this term. The local Doppler aberration term $\vnhat\cdot\vv_A$ is relatively large $\clo(10^{-3})$, but only contributes a dipole modulation and is not usually considered to be part of the cosmological signal since it depends on the local peculiar velocity.

In the background the angular diameter distance $\cld/2=\chi_*a_*$ is the Ricci focussing result for a flat FRW universe (Weyl is zero in the background). $\Delta_\gamma/4$ is the perturbation to this background result due to differences in time coordinate (hence perturbations in the expansion). The $\Phi$ term is the perturbation in the Ricci focussing to the background recombination time, and the sum $\Delta_\gamma/4-\Phi$ gives the total perturbation in observed size due to Ricci focussing. The decomposition is not gauge-invariant, but the sum $\zeta_\gamma\equiv \Delta_\gamma/4-\Phi$ is the gauge-invariant 3-curvature on constant photon density hypersurfaces.
In a sense the Ricci focussing perturbation is the ray bundle area analogue of the $\delta N$ formula for the comoving curvature perturbation from inflation: differences in expansion between different Hubble patches in inflation are seen as varying expansions of the ray bundle between the observer and different points on the last-scattering surface. An overdensity at recombination has larger local scale factor (described by the 3-curvature) than the reference point, so the beam cross-section shrinks compared to average as the ray bundle leaves the perturbation, leading to fixed physical sizes appearing smaller (see Fig.~\ref{diagram}).

The result for $\cld/2$ is valid on all scales, but the approximation of constant physical scales on the source plane may be modified by dynamics.
For the limiting case of super-horizon adiabatic modes where the physics is identical in single-field inflation,  $\frac{\Delta_\gamma}{4} -\Phi=\zeta$, where $\zeta$ is the conserved 3-curvature perturbation on equal (total) density hypersurfaces (in matter domination $\zeta \approx -5\Phi/3$). Note that overdensities have positive $\kappa$, positive $\Delta_\gamma$ and negative $\Phi$, so lensing convergence tends to cancel with the other terms: an overdensity has Ricci defocussing compared to average, but also acts as a magnifying lens, so the effects partly cancel. For a scale-invariant spectrum in matter domination and no ISW, in the flat-sky approximation $5/6$ of the effective magnification-temperature correlation is cancelled~\cite{Creminelli:2004pv,Creminelli:2011sq}, so it is important to consistently model both effects unless each one is separately negligible compared to the ISW signal. Note that a constant super-horizon $\Delta_\gamma$ perturbation appears to change $\cld$, however this is correct because it would be equivalent to observing a larger CMB background temperature, which for the same physics requires less time since recombination and hence corresponds to a smaller comoving radial distance to recombination.

The shear is significantly simpler since it vanishes in the background, and is given by the symmetric-trace free part of the Jacobi map $\cld_{\la IJ\ra} = a_*\chi_*\gamma_{IJ}$, where $\gamma_{IJ}$ is given in terms of the lensing potential as described above (see e.g.~\cite{Lewis:2006fu}). The shear is given solely by Weyl effects (lensing), so there are no complications with partial cancellations.

The shear and convergence are angular derivatives on the sky, and the lensing can be thought of as re-mapping points by a deflection angle (which is a conveniently simple effect to simulate). However, the Ricci focussing is not an angular derivative, and cannot be modelled using deflection angles: for non-trivial curvature of the recombination surface there is no local-isotropy preserving map onto the observed sphere. The spherical projection is intrinsically 3-dimensional, with the Ricci term being more like a radial deflection than an angular one.

\subsection{Sub-horizon dynamics}
\label{subhorizon}

Leading effects of dynamics may be important, because velocities come to dominate the CMB power spectrum at $l\sim 60$ (see Fig.~\ref{TTcontribs})  and so only the very lowest multipoles can safely be modelled as due to having only static super-horizon sources at recombination. Dynamics will modify the small physical scales observed, and hence lead to additional anisotropy in the distribution of the small-scale perturbations on the last-scattering surface. Here I give some simple arguments to quantify these effects, and hence determine a range of scales over which a simple squeezed analysis is likely to be robust.

Note that from the photon density perturbation evolution equation
\bea
\frac{\Delta_\gamma}{4} - \Phi &=&  \left[\frac{\Delta_\gamma}{4} - \Phi\right]_{0} -\frac{1}{3}\int_0^\eta \vgrad\cdot\vv_\gamma \,\d\eta'\nonumber
\\&=& \zeta_0 - \frac{1}{3}\int_0^\eta \vgrad\cdot\vv_\gamma \,\d\eta'
\eea
for adiabatic perturbations. We can therefore isolate the dynamical velocity compression at recombination by defining
\be
\delta N_v \equiv \clh \delta\eta_v \equiv \left[\frac{\Delta_\gamma}{4} - \Phi  -\zeta_0 \right],
\ee
which gives the fractional change in expansion due to the large-scale flows in the fluid compressing the small-scale perturbations, with $\delta \eta_v$ being the corresponding change in conformal time until recombination. Note that it is small on super-horizon scales since $\delta N_v \sim \clo((k/\clh)^2\zeta_0)$.
For scalar perturbations the flow is potential $\vv =\vgrad v_\gamma$, so there is a deflection angle model $\valpha_v=-\int\d\eta \vv_{\perp}/\chi_* = -\grad_{\vnhat} \int \d\eta v_\gamma/\chi_*^2$ for the dynamic effect of the bulk velocity-induced motion of small-scale perturbations in the plane of the sky, assuming the small perturbations co-move with the large-scale radiation flow. This gives rise to an effective shear and convergence compared to purely static evolution.
However, the 3-divergence in flow increases the local temperature; this is removed by additional expansion, so the total effect on the physical scale at the equal-temperature recombination surface is only the net effect.
Defining the angular velocity distortion potential
\be
\psi_v\equiv  - \int_0^{\eta_*} \frac{\d\eta v_\gamma}{\chi_*^2} = 3 \frac{\vgrad^{-2}\delta N_v }{\chi_*^2}
\ee
we can then calculate angular convergence $\kappa_v \equiv -\grad_I\grad^I \psi_v/2$ and shear $\gamma_{IJ}^v= \grad_{\la I}\grad_{J\ra}\psi_v$ with respect to primordial isotropic initial conditions, in exactly the same way as we do for lensing using the lensing potential.
The total effective fractional magnification and shear with respect to isotropy, including Ricci, Weyl and angular velocity dynamics, is then roughly
\be
\kappa^\eff \approx  \kappa +\kappa_v - \zeta_0
\qquad
\gamma^\eff_{IJ} \approx  \gamma_{IJ} +\gamma_{IJ}^v.
\label{kappa_eff}
\ee
In addition to the bulk movement of the small scale-perturbations, there will also be a change in their correlation length, determined roughly by the sound horizon. The fractional change in sound horizon due to recombination being delayed in a hotspot gives a scale magnification $\sim \delta\eta_v/\eta_*$
which reduces the effect of the velocity convergence but does not alter the shear. This is only a rough order of magnitude as there are of course many other effects, for example due to ionization fraction perturbations on the constant temperature slices (c.f.~\cite{Khatri:2009ja}).  There will be additional velocity effects from the radial components of the velocity, however these are more like power modulations similar to $\fnl$, and are therefore not potentially enhanced by $\d\ln C_l/{\d\ln l}$ due to the acoustic structure, and should be safely negligible at least as a contaminant of the large-scale lensing cross-correlation. There will also be an $\fnl$-like signal from modulation in power due to changes in the Silk scale and recombination thickness.

Figure~\ref{correlations} shows a numerical calculation of angular correlation power spectrum of the various contributions with the CMB temperature using \CAMB~\cite{Lewis:1999bs}. The effective magnification correlation is enhanced on large scales because there is a near-cancellation between convergence from large-scale lenses and the Ricci focussing $\zeta_0$ term; the large-scale modes contribute to the correlation with opposite sign to the ISW signal, so the signal is enhanced when most of the large-scale modes are cancelled by the $\zeta_0$ modulation.

 Note that temperature anisotropy directly probes the velocity via the Doppler effect which is $\clo(k\eta_*)$, but flow-generated convergence and shear depend on the effect of gradients of the velocities over time and hence are $\clo((k\eta_*)^2)$. It is therefore required to model the evolution and Doppler contributions to the large-scale temperature sources on significantly larger scales than is required for dynamical convergence and shear, or any other effect that depends on changes in time or temperature due to fluid compression. Furthermore the Doppler contribution to $\Delta T$ is uncorrelated to scalar densities, so the correlation of $\Delta T$ with dynamical compressions is small around $l\sim 60$. Only on smaller scales, roughly $l\agt 100$ does $\Delta T$ start to become dominated by the flow-induced density perturbations, and at that point a full analysis would be required to get an accurate result for the correlation.

These arguments and the numerical estimates indicate that a squeezed analysis neglecting second-order dynamics is likely to be accurate at $l\alt 100$. The $\clo((k\eta)^2)$ scaling of sub-horizon effects should ensure that the scaling of the full result is similar to the approximations here, and hence that this conclusion is fairly robust.
 Since $l\alt 100$ contains most of the ISW signal this is likely to be a good data cut for parameter studies that want to be confident of avoiding issues with sub-horizon dynamics. For primordial local non-Gaussianity the signal goes to zero at a modulation scale $l\approx 60$ where $C_l^{T\zeta_0^*}$ changes sign, so a restriction to $l\alt 60$ should ensure dynamical effects are small while losing little signal (for Planck $\sim 90\%$ of the temperature signal is at $l<60$, see e.g. Ref.~\cite[Fig.11 ]{Lewis:2011fk}). A cut at $l\alt 100$ is also sufficient for corrections due to finite thickness of the last scattering surface not to be large.

\section{The CMB bispectrum}

If we observe a small patch of the CMB, the statistics of the temperature fluctuations may vary as a function of position on the sky due to large-scale magnification, shear, and temperature modulation. Correlation of these modulations with the CMB temperature gives a bispectrum. If the (non-linear) modulated temperature is $\tT_{lm}$, for $l_1 < l_2,l_3$ and neglecting tiny non-linear effects on the large-scale mode (the linear short-leg approximation), the bispectrum from a set of Gaussian modulation fields $X_i$ is given by
\be
\la \tT_{l_1m_1} \tT_{l_2m_2} \tT_{l_3m_3} \ra \approx C_{l_1}^{TX_i} \left\la \frac{\delta}{\delta X_{i,l_1m_1}^*} \left(\tT_{l_2 m_2} \tT_{l_3 m_3}\right) \right\ra.
\label{squeezed_nonpert}
\ee
See Ref.~\cite{Lewis:2011au} for a derivation of this non-perturbative weakly-squeezed version of a well-known result: the bispectrum depends on the large-scale correlation of the modulation with the temperature, and the response of the small-scale power to changes in the large-scale modulation modes. Only the lensing effect is significantly large, so the non-perturbative average is only required over temperature and lensing potential modes. This gives results that (approximately) involve the lensed small-scale power spectra rather than the unlensed ones.


\subsection{Weyl lensing bispectrum}
The largest term is the lensing bispectrum, which is well understood~\cite{Seljak:1998nu,Zaldarriaga:2000ud,Hu:2000ee,Hanson:2009kg,Lewis:2011fk}. The temperature bispectrum signal is limited for detection by cosmic variance at about $5\sigma$, and should soon be detected by Planck at around 4$\sigma$. From Ref.~\cite{Lewis:2011fk} the reduced bispectrum is given to good accuracy on all scales by
\begin{multline}
b_{l_1 l_2 l_3} = \frac{1}{2}\left[(l_1(l_1+1) + l_2(l_2+1)-l_3(l_3+1)\right]C_{l_1}^{T\psi}\tC_{l_2}
\\ + \text{perms}
\label{fulllensing}
\end{multline}
where $\psi$ is the lensing potential. Note that this is non-perturbatively accurate in the small-scale modes, since the bispectrum involves
the \emph{lensed} small-scale power spectra -- an effect that would be missing in any complete purely second-order-only analysis.
In the flat-sky ultra-squeezed limit $l_1\ll l_2,l_3$ this reduces up to $\clo((l_1/l)^2)$ corrections to
\be
b_{l_1 l_2 l_3} \approx C_{l_1}^{T\kappa}\left[ \frac{1}{l^2} \frac{\d(l^2 \tilde{C}_l)}{d\ln l} + \cos 2\phi_{l_1l} \frac{\ud \tC_{l}}{\d\ln l}\right],
\label{lensingsqueezed}
\ee
where $\vl \equiv (\vl_2-\vl_3)/2$ and $\phi_{l_1 l}$ is the angle between $\vl_1$ and $\vl$ (the full sky version of this limit is given in Ref.~\cite[appendix B]{Pearson:2012ba}). The first term describes the isotropic squeezed bispectrum from magnification, and the second term the anisotropic quadrupolar bispectrum from shear. In the squeezed limit the two terms are effectively orthogonal. If neglected the temperature lensing bispectrum on its own is known to bias local primordial $\fnl$ estimators by $\fnl\sim 6$ for Planck ($\fnl\sim 9$ in the noise-free limit)~\cite{Smith:2006ud,Serra:2008wc,Hanson:2009kg}, but the distinctive shape makes lensing easily distinguishable.

Dynamical potential flow of small-scale perturbations gives an effective convergence and shear as described in Sec.~\ref{subhorizon}, and the form of the induced bispectrum from these effects will be identical to lensing.
The local $\vnhat\cdot\vv_A$ Doppler aberration in Eq.~\eqref{fulljabobi} also contributes the same way as dipolar convergence (there is no shear dipole)~\cite{Lewis:2006fu}, and hence gives a bispectrum of the form of Eq.~\eqref{fulllensing}. Since $\vv_A$ also determines the direction of the CMB temperature dipole, in a sense there is a large local $b_{1 l_2 l_3}$ bispectrum that should be detectable by Planck~\cite{Challinor:2002zh}, though this is not usually regarded as cosmological as it depends on the local velocity.

\subsection{Ricci focussing bispectrum}
\label{ricci}

The Ricci focussing $\zeta_0$ term gives an effective observed magnification of the observed angular scale (but no shear). In a conformal picture the effect of $\zeta\approx \Delta_\gamma/4 -\Phi$ in Eq.~\eqref{fulljabobi} is equivalent to the effect of distance perturbations when looking through an unperturbed universe to a last-scattering surface at equal time, with $\cld/2a_* = \chi_*(1 + \zeta_0)$. In this conformal picture the change in angular scale is coming purely from changes to the effective distance away\footnote{Note that in the equivalent conformal picture overdensities are further away, which is opposite to the expanding picture where overdensities recombine later and are actually slightly closer in the absence of large time delays.}, and the reduced bispectrum from the $\zeta_0$ distance modulation is then easily calculated to be
\be
b_{l_1 l_2 l_3} \approx C_{l_1}^{T\zeta_0^*}\frac{1}{2}\frac{\ud}{\ud\ln \chi_*}\left[\tC_{l_2}+\tC_{l_3}\right].
\label{RicciFull}
\ee
For a scale-invariant $C_l$ spectrum with sharp recombination $C_l$ is independent of $\chi_*$, so there is no visible scale modulation and hence no bispectrum, as expected; however in reality $C_l$ has significant acoustic oscillations. For the ultra-squeezed limit with $l_1\ll 200$, taking $l(l+1)C_l \propto f(l/\chi_*)$ this reduces approximately to
\be
b_{l_1 l_2 l_3} \approx -C_{l_1}^{T\zeta_0^*}\frac{1}{l^2}\frac{\ud}{\ud\ln l}(l^2 \tC_l),
\label{superhorizonsqueezed}
\ee
which as expected has the same form as the isotropic convergence term in Eq.~\eqref{lensingsqueezed}. Note that $C_{l_1}^{T\zeta_0^*}$ goes through zero at $l_1\sim 60$, so it should not be approximated with the $l_1\ll 60$ Sachs-Wolfe result.

Eq.~\eqref{superhorizonsqueezed} is the CMB equivalent of the $\sim \clp(k_1)\frac{\ud \clp(k)}{\ud\ln k} $ `consistency relation' bispectrum generated by single-field inflation~\cite{Maldacena:2002vr,Creminelli:2011sq}. Note however that it is not correct to evaluate an effective $\fnl = \clo(n_s-1)$ in the primordial power spectrum and then use that in the result for physical small-scale power modulations (Eq.~\eqref{fnlsqueezed} below), the reason being that $\d/\d \ln k$ does not commute with the CMB transfer functions. Since the transfer functions are highly oscillatory the signal from Eq.~\eqref{superhorizonsqueezed} is much larger (though still small). If there were no other signals it would be (un)detectable\footnote{I calculate biases, correlations and errors using a minor modification of \CAMB's public full-sky bispectrum code~\cite{Lewis:2011fk}.} at $\sim 0.4\sigma$ (using $l_1<100$), and projects onto $\fnl$ as a bias of $\fnl \sim 1$. If neglected it also contributes to the lensing signal, corresponding to a bias of around $5\%$ in lensing bispectrum amplitude, corresponding to a bias of about  $0.2\sigma$. However it is also easily calculable so these small biases can easily be modelled or subtracted.

The correction to the squeezed trispectrum is safely small compared to lensing, equivalent to only $1$---$2\%$ on the $\kappa$ power spectrum on the very lowest multipoles --- well below cosmic variance.

\subsection{Redshift and power modulation bispectrum}

In the single-surface approximation, the perturbed recombination surface is equal temperature. However, we see temperature anisotropies because the photons get different total amounts of redshifting along different lines of sight. When we observe small-scale anisotropies,
the observed photon temperatures are also modulated by the large-scale variations in redshifting between different points on the last-scattering surface. Note that this is exactly the same effect that gives rise to the linear large-scale anisotropies (Appendix~\ref{linearCMB}), so the small-scale anisotropies are just modulated by the large-scale $1+\Delta T$. Allowing for the fact that the small-scale power is also lensed this gives a reduced bispectrum\footnote{Note that $C_{l_1}$ here includes the ISW effect: viewing small-scale perturbations through a large evolving potential well will give a large-scale modulation of the observed redshift due to all relevant large-scale effects including ISW (the authors of Ref.~\cite{Creminelli:2011sq} agree with this contrary to a statement in the text).}
\be
b_{l_1 l_2 l_3} \approx C_{l_1} \left(\tC_{l_2} + \tC_{l_3}\right).
\label{b_tempmod}
\ee
Local primordial non-Gaussianity (for example from multi-field inflation) also generates local physical modulations in the small-scale perturbations, conventionally parameterized as $\zeta =[1+(3\fnl/5)\zeta_g]\zeta_g$. The CMB bispectrum from this signal is well known (see e.g. \cite{Bartolo:2004if,Komatsu:2003iq,Komatsu:2010hc}), and the full result can be easily calculated using \CAMB.
To compare to the other weakly squeezed signals, we can take the limit where $l_1$ corresponds to a scale much larger than the thickness of the last-scattering surface (e.g. the ultra-squeezed limit). The bispectrum then reduces to
\be
b_{l_1 l_2 l_3} \approx \frac{6}{5}\fnl C_{l_1}^{T\zeta_0^*}(\tC_{l_2} + \tC_{l_3}) ,
\label{fnlsqueezed}
\ee
where $\zeta_0^*$ is the primordial curvature perturbation evaluated at the position on the last-scattering surface (a numerical calculation of $C_l^{T\zeta_0^*}$ is included in Fig.~\ref{correlations}). This approximation is not accurate at $l_1\agt 60$ due to finite recombination thickness, but is nonetheless $\sim 90\%$ correlated with the full result and is accurate for low $l_1$. Note that it also trivially includes the effect of lensing on the primordial non-Gaussianity because the modulation affects the lensed small-scale power spectrum, see Ref.~\cite{Pearson:2012ba}.

On very large scales, neglecting ISW, $\Delta T\sim -\zeta_0^*/5$ so in the ultra-squeezed limit $b_{l_1 l_2 l_3}\sim -12\fnl C_{l_1}\tC_{l}$; hence the
always-present signal of Eq.~\eqref{b_tempmod} is comparable to $\fnl\sim -1/6$ and therefore negligible compared to cosmic variance limits of $\fnl\sim 2$ ~\cite{Creminelli:2011sq}. The shape of Eq.~\eqref{b_tempmod} is actually different to local non-Gaussianity because $C_{l_1}^{T\zeta_0^*}$ has opposite sign to $C_{l_1}$ for $l_1 \agt 60$, but remains even less detectable than the Ricci focussing effect because it is not enhanced by oscillations due to $\d \ln C_l/\d \ln l$ factors (which have amplitude $\sim 6$). Note however that the local dipole part of the temperature modulation gives an easily detectable dipolar power modulation signal~\cite{Challinor:2002zh}, which must be carefully subtracted to get at a primordial $\taunl$ trispectrum~\cite{Pearson:2012ba} (along with the dipole aberration discussed in the previous subsection).

\subsection{Full squeezed bispectrum}

\begin{table}
\centering
\begin{tabular} {| c || c || c | c | c || c|}
\hline
  \multicolumn{2}{|c||}{} &  \multicolumn{4}{|c|}{bias on $\fnl$}  \\ [0.5ex]
\hline
Data used & $\sigma_{\fnl}$ &  Weyl  &  Ricci  &  Redshift & Total \\ [0.5ex]
\hline
\hline
T                    & 4.3 & 9.5  & 1.5 &  -0.22  & 10.7  \\
Planck T             & 5.9 & 6.4  & 1.0 &  -0.22  & 7.1\\
\hline
T ($l_1 < 60$)       & 4.6 & 10.6 & 1.7 &  -0.25  & 12.0 \\
Planck T ($l_1 < 60$)& 6.2 & 7.0  & 1.1 &  -0.25  & 7.9 \\
\hline
T+E                    & 2.1 & 2.6  & 1.1 &  -0.05  & 3.7  \\
Planck T+E             & 5.2 & 4.3  & 1.0 &  -0.15  & 5.2\\
\hline
\end{tabular}

\caption{
Individual and total biases on primordial local-model
non-Gaussianity parameterized by $\fnl$ for CMB temperature and $E$-polarization data with Planck-like noise
(assuming isotropic coverage over the full sky with
sensitivity  $\Delta T  = \Delta Q/2=\Delta U/2 = 50\,\mu\text{K\,arcmin}$ [$N_l^{\temp}=N_l^E/4=2\times 10^{-4}\muK^2$]
and a beam FWHM of $7\, \rm{arcmin}$) or cosmic-variance limited data
with $\lmax=2000$. Results are assuming that non-$\fnl$ contributions are only significant at $l_1\le 300$ and negligible dynamical effects; the $l_1<60$ results are filtered to only use large scale modulations and are therefore immune to small-scale modulation effects.
The bias is the systematic error on $\fnl$ if the given contribution is neglected, which can be compared to $\sigma_{\fnl}$ which is the Fisher error estimate (including lensing signal variance).
}
\label{table:biases}
\end{table}

Even if there is no primordial non-Gaussianity, there is still a detectable CMB squeezed bispectrum due to the effect of perturbations on light propagation. This signal is all that is expected in single-field inflation models, and must be modelled to constrain primordial non-Gaussianity. The full signal is given by summing the three terms calculated above. The resulting projections onto primordial $\fnl$ are shown in Table~\ref{table:biases}. The Ricci focussing term also affects the Weyl lensing bispectrum, corresponding to a bias of $0.04$ for both Planck ($\sigma_{\rm{lens}}=0.26$) and cosmic variance ($\sigma_{\rm{lens}}=0.19$): a  bias of $\sim 0.2\sigma$.

Including all the terms the projection on $\fnl$ increases from $\fnl\sim 6.4$ for lensing only to a combined signal of $\fnl \sim 7.1$ for a Planck-like temperature measurement with the same assumptions as Ref.~\cite{Lewis:2011fk}. The increase is due to the cancellation between the large-scale Ricci focussing and Weyl convergence, which acts to \emph{increase} the total magnification-like signal compared to that calculated from lensing deflection angles alone. This is because the large-scale modes contribute to the temperature correlation with opposite sign to ISW. The correction is however small compared to the expected error bar of $\sigma_{\fnl} \sim 6$. The projection onto the lensing bispectrum is somewhat larger because Ricci focussing looks like convergence, and the potential bias is around $0.2\sigma$.

This is assuming that there are no dynamical sub-horizon effects that project onto $\fnl$; a conservative analysis can restrict to $l_1<60$ where the squeezed results should be robust, and dynamical and finite recombination-thickness lensing effects are small. For Planck this restriction only increases the error bar by $\sim 5\%$ (conversely any dynamical effects would have to be very large to significantly bias a full analysis, since only $\sim 10\%$ of the temperature signal is from higher $l_1$).

\section{Temperature-lensing correlation}

\begin{figure}
\includegraphics[width=8.7cm]{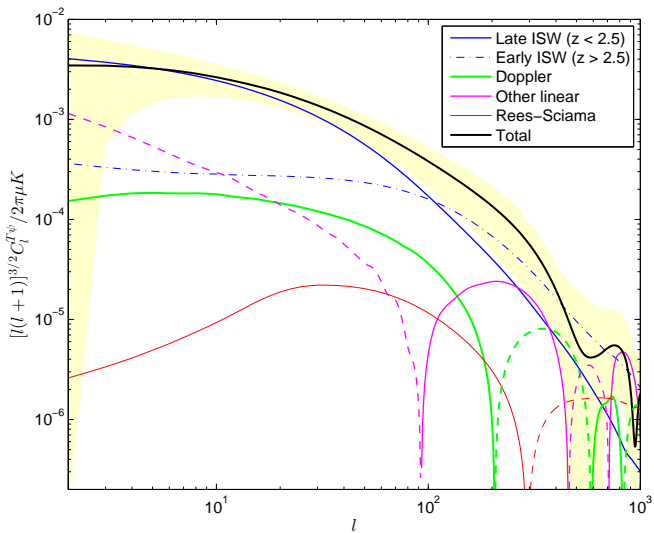}
\caption{Contributions to the CMB lensing deflection angle-temperature cross-correlation power spectrum, defining the lensing potential using a
recombination visibility weight function. Dashed lines are negative and the shaded band gives an estimate of cosmic variance sensitivity as in Fig.~\ref{correlations}.
The results are not directly observable on scales $l\gg 100$ where recombination surface effects become important (and hence acoustic oscillations), but indicate the order of magnitude of the complicating effects. At $l\alt 100$ where the signal is under good control it is dominated by early and late ISW, with a negative contribution from very large-scale modes at the lowest $l$ and a small Doppler contribution. The non-linear Rees-Sciama signal is always small compared to other effects.
The \CAMB\ Jan 2012 result using a delta-function visibility approximation and cutting sources very close to recombination agrees well with the black curve for $l\alt 100$. Note that this plot differs by a $\sqrt{l(l+1)}$ scaling factor compared to other plots in this paper.
\label{contribs}
}
\end{figure}

In the absence of primordial non-Gaussianity, the dominant contribution to the CMB bispectrum is CMB lensing. It is therefore important to model it accurately. Eq.~\eqref{fulllensing} is an accurate result for the lensing bispectrum, but requires a calculation of $C_l^{T\psi}$, the correlation between the temperature and the lensing potential. Here I investigate this signal in more detail to quantify the relevant physical contributions, assess where the calculation is reliable, and whether
non-linear Rees-Sciama contributions are important.

The following contributions may be relevant for the temperature lensing-potential correlation:
\begin{itemize}
\item Large-scale modes that span recombination and also act as lenses
\item The late Integrated Sachs Wolfe effect (late ISW) at low redshift from decaying potentials
\item The early Integrated Sachs Wolfe effect (early ISW) due to the transition from radiation to matter domination, and decaying modes
\item Lenses close to last-scattering being correlated to density perturbations that have infall giving a Doppler signal in the CMB
\item Doppler signal from scattering at reionization
\item Lenses at last-scattering that directly correlate perturbations to lensing at the recombination surface
\item Non-linear Rees-Sciama signal at low redshift from non-linear gravitational clustering
\item Non-linear SZ signal from scattering in clusters
\item Correlations due to foreground contaminants
\end{itemize}
The signal of most interest for parameter studies is the linear late ISW since it can be used as a constraint on dark energy models~\cite{Verde:2002mu} --- the CMB lensing potential happens to correlate at the $90\%$ level with the ISW~\cite{Mead:2010bv}. However the total temperature correlation drops off rapidly with scale as the ISW component of the signal tends to cancel between multiple perturbations along the line of sight. Except for the last three non-linear and foreground signals, all of the other effects are self-consistently calculated in linear theory by using a full standard line-of-sight Boltzmann code, and are included \CAMB's standard calculation\footnote{\CAMB\ Jan 2012 uses a single source-plane approximation and cuts sources very close to last-scattering; the result agrees well with the total result shown here at $l\alt 100$ but does not include most of the recombination surface effects that cannot be modelled consistently anyway.}. The thermal SZ signal is frequency dependent, but depending on frequency potentially important at $l\sim 100$~\cite{Munshi:2009fr}; I shall not consider this signal further here since it is zero at the SZ null and in principle is isolable by its frequency dependence, as are other potentially significant foreground signals like the correlation with the cosmic infrared background (CIB)~\cite{Song:2002sg,Ade:2011ap} (the CIB signal is also small at frequencies usually used for cosmological analysis).

The late-time ISW effect can also be calculated accurately on intermediate and small scales using the Limber approximation result~\cite{Verde:2002mu}
\be
l^3 C_l^{T\psi} \approx -4\pi^2\int_0^{\chi_*} \d\chi \frac{ \chi_*-\chi}{\chi^*} \frac{\partial}{\partial\eta} \clp_{\Phi}(k=l/\chi,\eta),
\ee
where $\clp_\Phi$ is the power spectrum of the potentials that can easily be calculated from the standard matter power spectrum.
The approximation of using $\partial_\eta \clp_\Phi$ is not strictly valid for non-linear evolution, but has been tested on simulations and appears to be adequate~\cite{Nishizawa:2007pv}, in which case the same result can also be used to calculate the full late ISW+Rees-Sciama result simply by using the non-linear power spectrum. I use the Halofit~\cite{Smith:2002dz} non-linear matter power spectrum model to estimate the difference in the total matter power spectrum due to non-linear evolution as a function of redshift. The non-linear Rees-Sciama signal has opposite sign to the ISW on small scales where structures are growing more rapidly than in linear theory.

For the purpose of estimating the size of effects close to the recombination surface I define the lensing potential using a visibility source weighting.
For the late-time contribution this gives results very close to those from using a single lens-plane approximation since the thickness of last-scattering is very small compared to its distance away from us. Using the smooth visibility weighting allows the importance of near and through-recombination contributions to be assessed, though the result is no longer directly related to what is observed and the non-lensing effects discussed in the previous section may also become comparable.

Numerical results are shown in Fig.~\ref{contribs}, demonstrating that even on large scales it is important to include both early and late ISW effects to avoid $1\sigma$ biases, as well as large-scale modes and a small correction from the Doppler signal.
The importance of Rees-Sciama is significantly smaller than claimed in Refs.~\cite{Verde:2002mu,Giovi:2004te,Mangilli:2009dr} and is always a small correction that is safe to ignore in the region where the signal is well understood\footnote{The disagreement arises even at the level of the linear power spectrum, where Ref.~\cite{Mangilli:2009dr} underestimate the lensing correlation with linear late-ISW power at $l\agt 100$ by a factor of $\clo(20)$. As this paper was being finalized Ref.~\cite{Junk:2012qt} appeared with results that agree with those here.}.

\section{Polarization correlation and bispectrum}

\begin{figure}
\includegraphics[width=8.7cm]{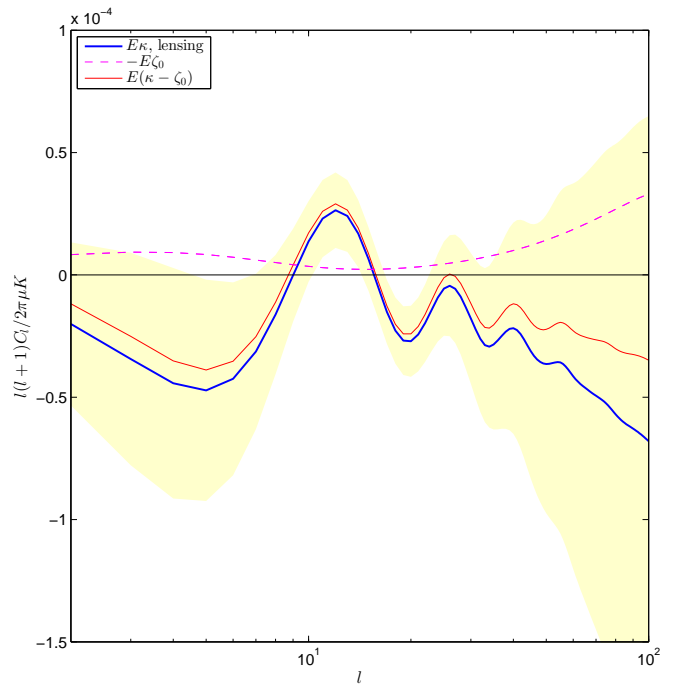}
\caption{Correlation of the large-scale $E$-polarization with the Weyl lensing convergence (which determines the shear), the Ricci focussing source $-\zeta_0^*$, and the sum which determines the total magnification correlation. The large-scale $E$ polarization is dominated by reionization scattering (optical depth $\tau=0.09$ here), and the correlation with the curvature perturbation at recombination is small. The shaded band shows an indicator of $\pm1\sigma$ sensitivity to systematic errors relative to the cosmic variance limit from lensing, given by $\pm \sqrt{ C_l^{EE} C^{\kappa\kappa}_l + (C_l^{E\kappa})^2}/l$, which is much smaller that the actual sensitivity with Planck due to significant large-scale $E$ polarization and lensing reconstruction noise.
\label{Ecorrelation}
}
\end{figure}


The extension to bispectra involving CMB polarization modes is straightforward, giving a Ricci focussing term of exactly the same form as Eq.~\eqref{RicciFull}:
 \be
b_{l_1 l_2 l_3}^{ijk} \approx C_{l_1}^{X_i\zeta_0^*}\frac{1}{2}\frac{\ud}{\ud\ln \chi_*}\left[\tC_{l_2}^{X_jX_k}+\tC_{l_3}^{X_jX_k}\right],
\label{RicciFullPol}
\ee
where $X_i,X_j,X_k$ are either $T$ or $E$.  There are also (larger) standard terms of the Weyl lensing polarization bispectrum~\cite{Hu:2000ee,Lewis:2011fk} that for brevity I shall not reproduce here. Anisotropic redshifting does not affect polarization components of the bispectra, giving
\be
b_{l_1 l_2 l_3}^{ijk} \approx C_{l_1}^{X_i T} \left(\delta_{kT} \tC_{l_2}^{X_j T} + \delta_{jT}\tC_{l_3}^{X_k T}\right).
\ee

The very large-scale polarization comes mostly from reionization, and is correlated with the lensing potential at up to $30\%$ which gives a Weyl lensing bispectrum contribution that is non-negligible at $l_1\alt 40$~\cite{Lewis:2011fk}.
There is also a Ricci focussing bispectrum from correlation of the curvature perturbations at recombination: see Fig.~\ref{Ecorrelation}.
Using polarization the lensing potential can be reconstructed with much higher signal to noise than with just temperature data, giving constraints on the cross-correlation with the temperature and polarization that are nearly signal-variance limited and a detection at about $8\sigma$.  Neglecting Ricci focussing to $\lmax=2000$ could then bias the total lensing bispectrum amplitude by $\sim 0.5\sigma$.

Large-scale polarization is especially useful around $l_1\sim 60$ where the $\fnl$ signal in the temperature goes to zero, and for Planck both contributions are small compared to errors there. It is harder to be sure of the bias on $\fnl$ from higher $l_1$ without having a full calculation of recombination level effects, since a more substantial portion of the $\fnl$ signal comes from $l_1\gg 100$, especially in the noise-free case. However if we assume that Ricci focussing and lensing are only important at $l_1<300$, for noise free $T$ and $E$ polarization data the Ricci focussing bispectrum gives a bias of $\fnl\sim 1$, corresponding to a $0.5\sigma$ bias on $\fnl$ in this case ($\sigma_{\fnl}\approx 2$). See Table.~\ref{table:biases} for a summary. The lensing and Ricci focussing signal between $100\le l_1 \le 300$ only changes the $\fnl$ bias by $\sim 0.2$, but dynamical and recombination surface effects from smaller scales could conceivably contribute more.

\section{Conclusions}

Even in the absence of primordial non-Gaussianity, the presence of perturbations on a wide range of scales inevitably leads to a non-zero bispectrum in the CMB.
By directly relating physical scales to observed angles, the squeezed limit can be robustly estimated using combinations of linear-theory results, giving results that agree with Refs.~\cite{Bartolo:2011wb,Creminelli:2011sq} in the appropriate limit. The signal is dominated by lensing though other terms can give small biases to lensing results at the $0.2\sigma$ level if ignored, up to $0.5\sigma$ with low-noise polarization data. I argued that the full-sky approximations presented here should be reliable compared to errors for modulation scales $l_1\alt 100$. This is sufficient for parameter estimation from the lensing signal, and local primordial local non-Gaussianity estimators can be filtered to include only the signal at low $l_1\alt 60$ with little loss of information for Planck.

The largest contribution to the bispectrum is from lensing-temperature correlations, which include important contributions from early as well as late ISW. On small scales the ISW signal becomes small, and a fully self-consistent calculation of all relevant recombination effects would be required to calculate the bispectrum. It remains possible that at $l_1\gg 100$ recombination-level correlations (e.g. between Doppler temperature anisotropies and lenses just in front of last scattering) generate detectable net squeezed non-Gaussianities, and this should be the subject of future numerical work. The same applies for other shapes of non-Gaussianity; only squeezed shapes have significant signal to noise involving a super-horizon mode, so unfortunately calculating contributions to other shapes requires detailed study of the sub-horizon regime.

Corrections to the lensing trispectrum (the lensing potential power spectrum) should be small compared to cosmic variance since the late-time lensing signal is much larger. However if the reconstruction is for B-mode lensing cleaning~\cite{Hirata:2003ka} with high signal to noise, so that the actual \emph{realization} of the reconstructed lensing potential matters, small errors due to Ricci focussing and dynamical anisotropies at recombination may require more detailed study.

\section{Acknowledgements}
 I thank Duncan Hanson for independently checking the Rees-Sciama result, Eiichiro Komatsu for suggesting the table, David Seery and Filippo Vernizzi for discussion and communication, Anna Mangilli for sharing numerical results of Ref.~\cite{Mangilli:2009dr}, and Ruth Pearson for spotting typos.
I acknowledge support from the Science and Technology Facilities Council [grant number ST/I000976/1].

\appendix

\section{Linear temperature anisotropies from anisotropic redshifting}
\label{linearCMB}

In the delta-function visibility approximation the equal-temperature recombination surface is observed by us to have anisotropies due to differing amounts of redshift in different directions. The result is very well known (for a more detailed review see e.g~\cite{Challinor:2004bd}), but I recap here for completeness and rephrase in a somewhat unfamiliar way to demonstrate the close analogy with the ray bundle focussing result.

In the conformal Newtonian Gauge, integrating the photon geodesic equation gives the energy today $E_A$ in terms at the energy at a (conformal) time $\eta$
\be
a_A E_A = a(\eta)E(\eta) \left[1 + \Psi(\eta) - \Psi_A  + \int_\eta^{\eta_A} \d\eta \partial_\eta (\Psi+\Phi)\right],
\label{pert_redshift}
\ee
where energies are those observed by observers with no (Newtonian gauge) peculiar velocity and the ISW integral is along the line of sight.
The potentials at $\eta$ and $A$ appear local, but arise from integrating total derivative terms along the photon path.
This can equivalently be written
\be
a_A E_A = a(\eta)E(\eta) \left[1 - \Phi(\eta) + \Phi_A  + \int_\eta^{\eta_A} \d\eta \partial\chi (\Psi+\Phi)\right],
\ee
where $\chi$ is the comoving radial distance and now the integral term involves spatial gradients (and hence is not zero in matter domination).
For a constant temperature surface the time of emission is $\eta=\eta_*+\delta\eta$, with $\delta\eta = -\delta \rho_\gamma / \rho_\gamma' = \Delta_\gamma/(4\clh)$, so the scale factor at the perturbed emission time is $a(\eta) = a_*(1+\Delta_\gamma/4)$.
Accounting for Doppler shifts due to peculiar velocities and converting to the fractional temperature perturbation then gives the temperature anisotropy
\bea
\Delta T(\vnhat)\! &=&
 \!\frac{\Delta_\gamma}{4} - \Phi + \vnhat\cdot (\vv_A-\vv)  + \int_\eta^{\eta_A} \!\d\eta \,\vnhat\cdot\vgrad (\Psi+\Phi)
 \nonumber
\\\!
&=& \zeta_\gamma + \vnhat\cdot (\vv_A-\vv)  + \int_\eta^{\eta_A} \!\d\eta \,\vnhat\cdot\vgrad (\Psi+\Phi)
\eea
where I defined $a_A(1-\Phi_A)=1$ and $\zeta_\gamma$ is the gauge-invariant 3-curvature on the constant-temperature recombination hypersurface.
On large scales there is a cancellation between $\zeta_\gamma$ which is positive in an overdensity, describing getting relatively hotter due to the Ricci defocussing leaving an overdensity (relative space contraction, so gets a relative blue shift), and the effect of climbing out the potential well: a line-of-sight integral over the line-of-sight gradients in the potentials. The net effect in matter domination is the Sachs-Wolfe result $\Phi/3$.

The form of this equation is very similar to the result for the beam area, Eq.~\eqref{fulljabobi}, involving the curvature perturbation at the source, and an integral of potential gradients (the radial analogue of the transverse gradients that give rise to convergence and shear). That they both depend on $\zeta_\gamma$ should be unsurprising since photon energy and beam size are both affected by the expansion of space ($\zeta_\gamma = \delta N\equiv \delta \ln a$).

\providecommand{\aj}{Astron. J. }\providecommand{\apj}{Astrophys. J.
  }\providecommand{\apjl}{Astrophys. J.
  }\providecommand{\mnras}{MNRAS}\providecommand{\aap}{Astron. Astrophys.}

\end{document}

%% file: macros.tex
\renewcommand{\d}{\text{d}}
\newcommand{\grad}{\nabla}

\newcommand{\muK}{\mu\rm{K}}

\newcommand{\lmax}{l_{\text{max}}}

\newcommand{\fnl}{f_{\rm{NL}}}
\newcommand{\taunl}{\tau_{\rm{NL}}}

\providecommand{\CAMB}{\textsc{camb}}

\providecommand{\LCDM}{{$\rm{\Lambda CDM}$}}

\newcommand{\begm}{\begin{pmatrix}}
\newcommand{\enm}{\end{pmatrix}}

\newcommand\ba{\begin{eqnarray}}
\newcommand\ea{\end{eqnarray}}
\newcommand\bea{\begin{eqnarray}}
\newcommand\eea{\end{eqnarray}}

\newcommand\be{\begin{equation}}
\newcommand\ee{\end{equation}}

\newcommand{\valpha}{{\boldsymbol{\alpha}}}
\newcommand{\vgrad}{{\boldsymbol{\nabla}}}

\newcommand{\la}{\langle}
\newcommand{\ra}{\rangle}

\newcommand{\ud}{{\rm d}}

\newcommand{\boldvec}[1]{{{\mathbf{#1}}}}

\newcommand{\vl}{\boldvec{l}}

\newcommand{\vn}{\boldvec{n}}

\newcommand{\vv}{\boldvec{v}}

\newcommand{\cld}{\mathcal{D}}

\newcommand{\clh}{\mathcal{H}}

\newcommand{\clo}{\mathcal{O}}
\newcommand{\clp}{\mathcal{P}}

\newcommand{\vnhat}{\hat{\vn}}

%% file: TPhi.bbl
\begin{thebibliography}{43}
\expandafter\ifx\csname natexlab\endcsname\relax\def\natexlab#1{#1}\fi
\expandafter\ifx\csname bibnamefont\endcsname\relax
  \def\bibnamefont#1{#1}\fi
\expandafter\ifx\csname bibfnamefont\endcsname\relax
  \def\bibfnamefont#1{#1}\fi
\expandafter\ifx\csname citenamefont\endcsname\relax
  \def\citenamefont#1{#1}\fi
\expandafter\ifx\csname url\endcsname\relax
  \def\url#1{\texttt{#1}}\fi
\expandafter\ifx\csname urlprefix\endcsname\relax\def\urlprefix{URL }\fi
\providecommand{\bibinfo}[2]{#2}
\providecommand{\eprint}[2][]{\url{#2}}

\bibitem[{\citenamefont{Pitrou et~al.}(2010)\citenamefont{Pitrou, Uzan, and
  Bernardeau}}]{Pitrou:2010sn}
\bibinfo{author}{\bibfnamefont{C.}~\bibnamefont{Pitrou}},
  \bibinfo{author}{\bibfnamefont{J.-P.} \bibnamefont{Uzan}}, \bibnamefont{and}
  \bibinfo{author}{\bibfnamefont{F.}~\bibnamefont{Bernardeau}},
  \bibinfo{journal}{JCAP} \textbf{\bibinfo{volume}{1007}}, \bibinfo{pages}{003}
  (\bibinfo{year}{2010}), \eprint{1003.0481}.

\bibitem[{\citenamefont{Creminelli and
  Zaldarriaga}(2004{\natexlab{a}})}]{Creminelli:2004pv}
\bibinfo{author}{\bibfnamefont{P.}~\bibnamefont{Creminelli}} \bibnamefont{and}
  \bibinfo{author}{\bibfnamefont{M.}~\bibnamefont{Zaldarriaga}},
  \bibinfo{journal}{Phys. Rev.} \textbf{\bibinfo{volume}{D70}},
  \bibinfo{pages}{083532} (\bibinfo{year}{2004}{\natexlab{a}}),
  \eprint{astro-ph/0405428}.

\bibitem[{\citenamefont{Boubekeur et~al.}(2009)\citenamefont{Boubekeur,
  Creminelli, D'Amico, Norena, and Vernizzi}}]{Boubekeur:2009uk}
\bibinfo{author}{\bibfnamefont{L.}~\bibnamefont{Boubekeur}},
  \bibinfo{author}{\bibfnamefont{P.}~\bibnamefont{Creminelli}},
  \bibinfo{author}{\bibfnamefont{G.}~\bibnamefont{D'Amico}},
  \bibinfo{author}{\bibfnamefont{J.}~\bibnamefont{Norena}}, \bibnamefont{and}
  \bibinfo{author}{\bibfnamefont{F.}~\bibnamefont{Vernizzi}},
  \bibinfo{journal}{JCAP} \textbf{\bibinfo{volume}{0908}}, \bibinfo{pages}{029}
  (\bibinfo{year}{2009}), \eprint{0906.0980}.

\bibitem[{\citenamefont{Bartolo et~al.}(2012)\citenamefont{Bartolo, Matarrese,
  and Riotto}}]{Bartolo:2011wb}
\bibinfo{author}{\bibfnamefont{N.}~\bibnamefont{Bartolo}},
  \bibinfo{author}{\bibfnamefont{S.}~\bibnamefont{Matarrese}},
  \bibnamefont{and} \bibinfo{author}{\bibfnamefont{A.}~\bibnamefont{Riotto}},
  \bibinfo{journal}{JCAP} \textbf{\bibinfo{volume}{1202}}, \bibinfo{pages}{017}
  (\bibinfo{year}{2012}), \eprint{1109.2043}.

\bibitem[{\citenamefont{Creminelli et~al.}(2011)\citenamefont{Creminelli,
  Pitrou, and Vernizzi}}]{Creminelli:2011sq}
\bibinfo{author}{\bibfnamefont{P.}~\bibnamefont{Creminelli}},
  \bibinfo{author}{\bibfnamefont{C.}~\bibnamefont{Pitrou}}, \bibnamefont{and}
  \bibinfo{author}{\bibfnamefont{F.}~\bibnamefont{Vernizzi}},
  \bibinfo{journal}{JCAP} \textbf{\bibinfo{volume}{1111}}, \bibinfo{pages}{025}
  (\bibinfo{year}{2011}), \eprint{1109.1822}.

\bibitem[{\citenamefont{Senatore and Zaldarriaga}(2012)}]{Senatore:2012nq}
\bibinfo{author}{\bibfnamefont{L.}~\bibnamefont{Senatore}} \bibnamefont{and}
  \bibinfo{author}{\bibfnamefont{M.}~\bibnamefont{Zaldarriaga}}
  (\bibinfo{year}{2012}), \eprint{1203.6354}.

\bibitem[{\citenamefont{Maldacena}(2003)}]{Maldacena:2002vr}
\bibinfo{author}{\bibfnamefont{J.~M.} \bibnamefont{Maldacena}},
  \bibinfo{journal}{JHEP} \textbf{\bibinfo{volume}{05}}, \bibinfo{pages}{013}
  (\bibinfo{year}{2003}), \eprint{astro-ph/0210603}.

\bibitem[{\citenamefont{Okamoto and Hu}(2003)}]{Okamoto03}
\bibinfo{author}{\bibfnamefont{T.}~\bibnamefont{Okamoto}} \bibnamefont{and}
  \bibinfo{author}{\bibfnamefont{W.}~\bibnamefont{Hu}}, \bibinfo{journal}{Phys.
  Rev.} \textbf{\bibinfo{volume}{D67}}, \bibinfo{pages}{083002}
  (\bibinfo{year}{2003}), \eprint{astro-ph/0301031}.

\bibitem[{\citenamefont{Hanson et~al.}(2010)\citenamefont{Hanson, Challinor,
  and Lewis}}]{Hanson:2009kr}
\bibinfo{author}{\bibfnamefont{D.}~\bibnamefont{Hanson}},
  \bibinfo{author}{\bibfnamefont{A.}~\bibnamefont{Challinor}},
  \bibnamefont{and} \bibinfo{author}{\bibfnamefont{A.}~\bibnamefont{Lewis}},
  \bibinfo{journal}{General Relativity and Gravitation}
  \textbf{\bibinfo{volume}{42}}, \bibinfo{pages}{2197} (\bibinfo{year}{2010}),
  \eprint{0911.0612}.

\bibitem[{\citenamefont{Challinor and Lewis}(2011)}]{Challinor:2011bk}
\bibinfo{author}{\bibfnamefont{A.}~\bibnamefont{Challinor}} \bibnamefont{and}
  \bibinfo{author}{\bibfnamefont{A.}~\bibnamefont{Lewis}},
  \bibinfo{journal}{Phys.Rev.} \textbf{\bibinfo{volume}{D84}},
  \bibinfo{pages}{043516} (\bibinfo{year}{2011}), \eprint{1105.5292}.

\bibitem[{\citenamefont{Schmidt and Jeong}(2012)}]{Schmidt:2012ne}
\bibinfo{author}{\bibfnamefont{F.}~\bibnamefont{Schmidt}} \bibnamefont{and}
  \bibinfo{author}{\bibfnamefont{D.}~\bibnamefont{Jeong}}
  (\bibinfo{year}{2012}), \eprint{1204.3625}.

\bibitem[{\citenamefont{Creminelli and
  Zaldarriaga}(2004{\natexlab{b}})}]{Creminelli:2004yq}
\bibinfo{author}{\bibfnamefont{P.}~\bibnamefont{Creminelli}} \bibnamefont{and}
  \bibinfo{author}{\bibfnamefont{M.}~\bibnamefont{Zaldarriaga}},
  \bibinfo{journal}{JCAP} \textbf{\bibinfo{volume}{0410}}, \bibinfo{pages}{006}
  (\bibinfo{year}{2004}{\natexlab{b}}), \eprint{astro-ph/0407059}.

\bibitem[{\citenamefont{Lewis and Challinor}(2006)}]{Lewis:2006fu}
\bibinfo{author}{\bibfnamefont{A.}~\bibnamefont{Lewis}} \bibnamefont{and}
  \bibinfo{author}{\bibfnamefont{A.}~\bibnamefont{Challinor}},
  \bibinfo{journal}{Phys. Rept.} \textbf{\bibinfo{volume}{429}},
  \bibinfo{pages}{1} (\bibinfo{year}{2006}), \eprint{astro-ph/0601594}.

\bibitem[{\citenamefont{Sachs}(1961)}]{Sachs61}
\bibinfo{author}{\bibfnamefont{R.}~\bibnamefont{Sachs}},
  \bibinfo{journal}{Proc. Roy. Soc. Lon.} \textbf{\bibinfo{volume}{264}},
  \bibinfo{pages}{309} (\bibinfo{year}{1961}).

\bibitem[{\citenamefont{Bernardeau et~al.}(2010)\citenamefont{Bernardeau,
  Bonvin, and Vernizzi}}]{Bernardeau:2009bm}
\bibinfo{author}{\bibfnamefont{F.}~\bibnamefont{Bernardeau}},
  \bibinfo{author}{\bibfnamefont{C.}~\bibnamefont{Bonvin}}, \bibnamefont{and}
  \bibinfo{author}{\bibfnamefont{F.}~\bibnamefont{Vernizzi}},
  \bibinfo{journal}{Phys. Rev.} \textbf{\bibinfo{volume}{D81}},
  \bibinfo{pages}{083002} (\bibinfo{year}{2010}), \eprint{0911.2244}.

\bibitem[{\citenamefont{Hu and Cooray}(2001)}]{Hu:2001yq}
\bibinfo{author}{\bibfnamefont{W.}~\bibnamefont{Hu}} \bibnamefont{and}
  \bibinfo{author}{\bibfnamefont{A.}~\bibnamefont{Cooray}},
  \bibinfo{journal}{Phys. Rev.} \textbf{\bibinfo{volume}{D63}},
  \bibinfo{pages}{023504} (\bibinfo{year}{2001}), \eprint{astro-ph/0008001}.

\bibitem[{\citenamefont{Khatri and Wandelt}(2010)}]{Khatri:2009ja}
\bibinfo{author}{\bibfnamefont{R.}~\bibnamefont{Khatri}} \bibnamefont{and}
  \bibinfo{author}{\bibfnamefont{B.~D.} \bibnamefont{Wandelt}},
  \bibinfo{journal}{Phys.Rev.} \textbf{\bibinfo{volume}{D81}},
  \bibinfo{pages}{103518} (\bibinfo{year}{2010}), \eprint{0903.0871}.

\bibitem[{\citenamefont{Lewis et~al.}(2000)\citenamefont{Lewis, Challinor, and
  Lasenby}}]{Lewis:1999bs}
\bibinfo{author}{\bibfnamefont{A.}~\bibnamefont{Lewis}},
  \bibinfo{author}{\bibfnamefont{A.}~\bibnamefont{Challinor}},
  \bibnamefont{and} \bibinfo{author}{\bibfnamefont{A.}~\bibnamefont{Lasenby}},
  \bibinfo{journal}{Astrophys. J.} \textbf{\bibinfo{volume}{538}},
  \bibinfo{pages}{473} (\bibinfo{year}{2000}), \eprint{astro-ph/9911177}.

\bibitem[{\citenamefont{Lewis et~al.}(2011)\citenamefont{Lewis, Challinor, and
  Hanson}}]{Lewis:2011fk}
\bibinfo{author}{\bibfnamefont{A.}~\bibnamefont{Lewis}},
  \bibinfo{author}{\bibfnamefont{A.}~\bibnamefont{Challinor}},
  \bibnamefont{and} \bibinfo{author}{\bibfnamefont{D.}~\bibnamefont{Hanson}},
  \bibinfo{journal}{JCAP} \textbf{\bibinfo{volume}{1103}}, \bibinfo{pages}{018}
  (\bibinfo{year}{2011}), \eprint{1101.2234}.

\bibitem[{\citenamefont{Lewis}(2011)}]{Lewis:2011au}
\bibinfo{author}{\bibfnamefont{A.}~\bibnamefont{Lewis}},
  \bibinfo{journal}{JCAP} \textbf{\bibinfo{volume}{1110}}, \bibinfo{pages}{026}
  (\bibinfo{year}{2011}), \eprint{1107.5431}.

\bibitem[{\citenamefont{Seljak and Zaldarriaga}(1999)}]{Seljak:1998nu}
\bibinfo{author}{\bibfnamefont{U.}~\bibnamefont{Seljak}} \bibnamefont{and}
  \bibinfo{author}{\bibfnamefont{M.}~\bibnamefont{Zaldarriaga}},
  \bibinfo{journal}{Phys. Rev.} \textbf{\bibinfo{volume}{D60}},
  \bibinfo{pages}{043504} (\bibinfo{year}{1999}), \eprint{astro-ph/9811123}.

\bibitem[{\citenamefont{Zaldarriaga}(2000)}]{Zaldarriaga:2000ud}
\bibinfo{author}{\bibfnamefont{M.}~\bibnamefont{Zaldarriaga}},
  \bibinfo{journal}{Phys. Rev.} \textbf{\bibinfo{volume}{D62}},
  \bibinfo{pages}{063510} (\bibinfo{year}{2000}), \eprint{astro-ph/9910498}.

\bibitem[{\citenamefont{Hu}(2000)}]{Hu:2000ee}
\bibinfo{author}{\bibfnamefont{W.}~\bibnamefont{Hu}}, \bibinfo{journal}{Phys.
  Rev.} \textbf{\bibinfo{volume}{D62}}, \bibinfo{pages}{043007}
  (\bibinfo{year}{2000}), \eprint{astro-ph/0001303}.

\bibitem[{\citenamefont{Hanson et~al.}(2009)\citenamefont{Hanson, Smith,
  Challinor, and Liguori}}]{Hanson:2009kg}
\bibinfo{author}{\bibfnamefont{D.}~\bibnamefont{Hanson}},
  \bibinfo{author}{\bibfnamefont{K.~M.} \bibnamefont{Smith}},
  \bibinfo{author}{\bibfnamefont{A.}~\bibnamefont{Challinor}},
  \bibnamefont{and} \bibinfo{author}{\bibfnamefont{M.}~\bibnamefont{Liguori}},
  \bibinfo{journal}{Phys. Rev.} \textbf{\bibinfo{volume}{D80}},
  \bibinfo{pages}{083004} (\bibinfo{year}{2009}), \eprint{0905.4732}.

\bibitem[{\citenamefont{Pearson et~al.}(2012)\citenamefont{Pearson, Lewis, and
  Regan}}]{Pearson:2012ba}
\bibinfo{author}{\bibfnamefont{R.}~\bibnamefont{Pearson}},
  \bibinfo{author}{\bibfnamefont{A.}~\bibnamefont{Lewis}}, \bibnamefont{and}
  \bibinfo{author}{\bibfnamefont{D.}~\bibnamefont{Regan}},
  \bibinfo{journal}{JCAP} \textbf{\bibinfo{volume}{1203}}, \bibinfo{pages}{011}
  (\bibinfo{year}{2012}), \eprint{1201.1010}.

\bibitem[{\citenamefont{Smith and Zaldarriaga}(2011)}]{Smith:2006ud}
\bibinfo{author}{\bibfnamefont{K.~M.} \bibnamefont{Smith}} \bibnamefont{and}
  \bibinfo{author}{\bibfnamefont{M.}~\bibnamefont{Zaldarriaga}},
  \bibinfo{journal}{Mon.Not.Roy.Astron.Soc.} \textbf{\bibinfo{volume}{417}},
  \bibinfo{pages}{2} (\bibinfo{year}{2011}), \eprint{astro-ph/0612571}.

\bibitem[{\citenamefont{Serra and Cooray}(2008)}]{Serra:2008wc}
\bibinfo{author}{\bibfnamefont{P.}~\bibnamefont{Serra}} \bibnamefont{and}
  \bibinfo{author}{\bibfnamefont{A.}~\bibnamefont{Cooray}},
  \bibinfo{journal}{Phys. Rev.} \textbf{\bibinfo{volume}{D77}},
  \bibinfo{pages}{107305} (\bibinfo{year}{2008}), \eprint{0801.3276}.

\bibitem[{\citenamefont{Challinor and van Leeuwen}(2002)}]{Challinor:2002zh}
\bibinfo{author}{\bibfnamefont{A.}~\bibnamefont{Challinor}} \bibnamefont{and}
  \bibinfo{author}{\bibfnamefont{F.}~\bibnamefont{van Leeuwen}},
  \bibinfo{journal}{Phys. Rev.} \textbf{\bibinfo{volume}{D65}},
  \bibinfo{pages}{103001} (\bibinfo{year}{2002}), \eprint{astro-ph/0112457}.

\bibitem[{\citenamefont{Bartolo et~al.}(2004)\citenamefont{Bartolo, Komatsu,
  Matarrese, and Riotto}}]{Bartolo:2004if}
\bibinfo{author}{\bibfnamefont{N.}~\bibnamefont{Bartolo}},
  \bibinfo{author}{\bibfnamefont{E.}~\bibnamefont{Komatsu}},
  \bibinfo{author}{\bibfnamefont{S.}~\bibnamefont{Matarrese}},
  \bibnamefont{and} \bibinfo{author}{\bibfnamefont{A.}~\bibnamefont{Riotto}},
  \bibinfo{journal}{Phys. Rept.} \textbf{\bibinfo{volume}{402}},
  \bibinfo{pages}{103} (\bibinfo{year}{2004}), \eprint{astro-ph/0406398}.

\bibitem[{\citenamefont{Komatsu et~al.}(2005)\citenamefont{Komatsu, Spergel,
  and Wandelt}}]{Komatsu:2003iq}
\bibinfo{author}{\bibfnamefont{E.}~\bibnamefont{Komatsu}},
  \bibinfo{author}{\bibfnamefont{D.~N.} \bibnamefont{Spergel}},
  \bibnamefont{and} \bibinfo{author}{\bibfnamefont{B.~D.}
  \bibnamefont{Wandelt}}, \bibinfo{journal}{Astrophys. J.}
  \textbf{\bibinfo{volume}{634}}, \bibinfo{pages}{14} (\bibinfo{year}{2005}),
  \eprint{astro-ph/0305189}.

\bibitem[{\citenamefont{Komatsu}(2010)}]{Komatsu:2010hc}
\bibinfo{author}{\bibfnamefont{E.}~\bibnamefont{Komatsu}},
  \bibinfo{journal}{Class. Quant. Grav.} \textbf{\bibinfo{volume}{27}},
  \bibinfo{pages}{124010} (\bibinfo{year}{2010}), \eprint{1003.6097}.

\bibitem[{\citenamefont{Verde and Spergel}(2002)}]{Verde:2002mu}
\bibinfo{author}{\bibfnamefont{L.}~\bibnamefont{Verde}} \bibnamefont{and}
  \bibinfo{author}{\bibfnamefont{D.~N.} \bibnamefont{Spergel}},
  \bibinfo{journal}{Phys. Rev.} \textbf{\bibinfo{volume}{D65}},
  \bibinfo{pages}{043007} (\bibinfo{year}{2002}), \eprint{astro-ph/0108179}.

\bibitem[{\citenamefont{Mead et~al.}(2011)\citenamefont{Mead, Lewis, and
  King}}]{Mead:2010bv}
\bibinfo{author}{\bibfnamefont{J.~M.~G.} \bibnamefont{Mead}},
  \bibinfo{author}{\bibfnamefont{A.}~\bibnamefont{Lewis}}, \bibnamefont{and}
  \bibinfo{author}{\bibfnamefont{L.~J.} \bibnamefont{King}},
  \bibinfo{journal}{Phys. Rev.} \textbf{\bibinfo{volume}{D83}},
  \bibinfo{pages}{023507} (\bibinfo{year}{2011}), \eprint{1009.1549}.

\bibitem[{\citenamefont{Munshi et~al.}(2011)\citenamefont{Munshi, Valageas,
  Cooray, and Heavens}}]{Munshi:2009fr}
\bibinfo{author}{\bibfnamefont{D.}~\bibnamefont{Munshi}},
  \bibinfo{author}{\bibfnamefont{P.}~\bibnamefont{Valageas}},
  \bibinfo{author}{\bibfnamefont{A.}~\bibnamefont{Cooray}}, \bibnamefont{and}
  \bibinfo{author}{\bibfnamefont{A.}~\bibnamefont{Heavens}},
  \bibinfo{journal}{Mon.Not.Roy.Astron.Soc.} \textbf{\bibinfo{volume}{414}},
  \bibinfo{pages}{3173} (\bibinfo{year}{2011}), \eprint{0907.3229}.

\bibitem[{\citenamefont{Song et~al.}(2003)\citenamefont{Song, Cooray, Knox, and
  Zaldarriaga}}]{Song:2002sg}
\bibinfo{author}{\bibfnamefont{Y.-S.} \bibnamefont{Song}},
  \bibinfo{author}{\bibfnamefont{A.}~\bibnamefont{Cooray}},
  \bibinfo{author}{\bibfnamefont{L.}~\bibnamefont{Knox}}, \bibnamefont{and}
  \bibinfo{author}{\bibfnamefont{M.}~\bibnamefont{Zaldarriaga}},
  \bibinfo{journal}{Astrophys. J.} \textbf{\bibinfo{volume}{590}},
  \bibinfo{pages}{664} (\bibinfo{year}{2003}), \eprint{astro-ph/0209001}.

\bibitem[{\citenamefont{Ade et~al.}(2011)}]{Ade:2011ap}
\bibinfo{author}{\bibfnamefont{P.}~\bibnamefont{Ade}} \bibnamefont{et~al.}
  (\bibinfo{collaboration}{Planck Collaboration}),
  \bibinfo{journal}{Astron.Astrophys.} \textbf{\bibinfo{volume}{536}},
  \bibinfo{pages}{A18} (\bibinfo{year}{2011}), \eprint{1101.2028}.

\bibitem[{\citenamefont{Nishizawa et~al.}(2008)\citenamefont{Nishizawa,
  Komatsu, Yoshida, Takahashi, and Sugiyama}}]{Nishizawa:2007pv}
\bibinfo{author}{\bibfnamefont{A.~J.} \bibnamefont{Nishizawa}},
  \bibinfo{author}{\bibfnamefont{E.}~\bibnamefont{Komatsu}},
  \bibinfo{author}{\bibfnamefont{N.}~\bibnamefont{Yoshida}},
  \bibinfo{author}{\bibfnamefont{R.}~\bibnamefont{Takahashi}},
  \bibnamefont{and} \bibinfo{author}{\bibfnamefont{N.}~\bibnamefont{Sugiyama}},
  \bibinfo{journal}{\apjl} \textbf{\bibinfo{volume}{676}}, \bibinfo{pages}{L93}
  (\bibinfo{year}{2008}), \eprint{0711.1696}.

\bibitem[{\citenamefont{Smith et~al.}(2003)}]{Smith:2002dz}
\bibinfo{author}{\bibfnamefont{R.~E.} \bibnamefont{Smith}} \bibnamefont{et~al.}
  (\bibinfo{collaboration}{The Virgo Consortium}), \bibinfo{journal}{Mon. Not.
  Roy. Astron. Soc.} \textbf{\bibinfo{volume}{341}}, \bibinfo{pages}{1311}
  (\bibinfo{year}{2003}), \eprint{astro-ph/0207664}.

\bibitem[{\citenamefont{Giovi et~al.}(2005)\citenamefont{Giovi, Baccigalupi,
  and Perrotta}}]{Giovi:2004te}
\bibinfo{author}{\bibfnamefont{F.}~\bibnamefont{Giovi}},
  \bibinfo{author}{\bibfnamefont{C.}~\bibnamefont{Baccigalupi}},
  \bibnamefont{and} \bibinfo{author}{\bibfnamefont{F.}~\bibnamefont{Perrotta}},
  \bibinfo{journal}{Phys. Rev.} \textbf{\bibinfo{volume}{D71}},
  \bibinfo{pages}{103009} (\bibinfo{year}{2005}), \eprint{astro-ph/0411702}.

\bibitem[{\citenamefont{Mangilli and Verde}(2009)}]{Mangilli:2009dr}
\bibinfo{author}{\bibfnamefont{A.}~\bibnamefont{Mangilli}} \bibnamefont{and}
  \bibinfo{author}{\bibfnamefont{L.}~\bibnamefont{Verde}},
  \bibinfo{journal}{Phys. Rev.} \textbf{\bibinfo{volume}{D80}},
  \bibinfo{pages}{123007} (\bibinfo{year}{2009}), \eprint{0906.2317}.

\bibitem[{\citenamefont{Junk and Komatsu}(2012)}]{Junk:2012qt}
\bibinfo{author}{\bibfnamefont{V.}~\bibnamefont{Junk}} \bibnamefont{and}
  \bibinfo{author}{\bibfnamefont{E.}~\bibnamefont{Komatsu}}
  (\bibinfo{year}{2012}), \eprint{1204.3789}.

\bibitem[{\citenamefont{Hirata and Seljak}(2003)}]{Hirata:2003ka}
\bibinfo{author}{\bibfnamefont{C.~M.} \bibnamefont{Hirata}} \bibnamefont{and}
  \bibinfo{author}{\bibfnamefont{U.}~\bibnamefont{Seljak}},
  \bibinfo{journal}{Phys. Rev.} \textbf{\bibinfo{volume}{D68}},
  \bibinfo{pages}{083002} (\bibinfo{year}{2003}), \eprint{astro-ph/0306354}.

\bibitem[{\citenamefont{Challinor}(2004)}]{Challinor:2004bd}
\bibinfo{author}{\bibfnamefont{A.}~\bibnamefont{Challinor}}
  (\bibinfo{year}{2004}), \eprint{astro-ph/0403344}.

\end{thebibliography}
